\documentclass[conference]{IEEEtran}
\IEEEoverridecommandlockouts
\usepackage[utf8]{inputenc}
\usepackage{cite}
\usepackage{cuted}
\usepackage{subcaption}
\usepackage{amsmath,amssymb,amsfonts}
\usepackage{bm}
\usepackage{cancel}
\usepackage[ruled,linesnumbered]{algorithm2e}
\usepackage{algorithmic}
\usepackage{graphicx}
\usepackage{textcomp}
\usepackage{xcolor}
\usepackage[export]{adjustbox}
\usepackage{xurl}
\usepackage{comment}
\usepackage{paralist} 
\usepackage{enumitem} 
\usepackage[T1]{fontenc}
\DeclareMathAlphabet{\mathpzc}{OT1}{pzc}{m}{it}
\usepackage{tabularx}
\usepackage{booktabs}
\usepackage{multirow}
\usepackage{mathtools}
\usepackage{makecell}
\usepackage{soul}
\usepackage{balance}
\usepackage{hhline}
\usepackage{booktabs}
\usepackage{enumitem}

\usepackage{comment}

\usepackage[most]{tcolorbox}

\usepackage{nomencl}
\makenomenclature
\usepackage{etoolbox}
\renewcommand\nomgroup[1]{%
  \ifstrequal{#1}{P}{\vspace{10pt}\item[\textbf{Parameters}]}{%
  \ifstrequal{#1}{V}{\vspace{10pt}\item[\textbf{Variables}]}{}}{%
  \ifstrequal{#1}{S}{\vspace{10pt}\item[\textbf{Sets}]}{}}%
}

\title{Spatiotemporal Impact Analysis of Hurricanes and Storm Surges on Power Systems}

\author{\IEEEauthorblockN{Abodh Poudyal$^\dagger$, Charlotte Wertz, Amy Mi Nguyen, \\Sajjad Uddin Mahmud, Anamika Dubey}
\IEEEauthorblockA{Washington State University\\
Pullman, Washington, USA\\
Email: $^\dagger$<abodh.poudyal@wsu.edu>
\vspace{-3em}}

\and
\IEEEauthorblockN{Vibha Gunturi}
\IEEEauthorblockA{The Charter School of Wilmington\\
 Wilmington, DE, USA\\
Email: gunturi.vibha@charterschool.org 
\thanks{This work was supported by National Science Foundation (NSF) CAREER Grant \#1944142.}
}
}

\begin{document}
\bstctlcite{IEEEexample:BSTcontrol}
\maketitle
\thispagestyle{plain}
\pagestyle{plain}

\begin{abstract}
This paper develops a spatiotemporal probabilistic impact assessment framework to analyze and quantify the compounding effect of hurricanes and storm surges on the bulk power grid. The probabilistic synthetic hurricane tracks are generated using historical hurricane data, and storm surge scenarios are generated based on observed hurricane parameters. The system losses are modeled using a loss metric that quantifies the total load loss. The overall simulation is performed on the synthetic Texas 2000-bus system mapped on the geographical footprint of Texas. The results show that power substation inundation due to storm surge creates additional load losses as the hurricane traverses inland. 

\end{abstract}

\vspace{-5pt}

\section{Introduction}
Hurricanes account for over a trillion dollars in economic losses and are considered to be the major reason for power outages in the US~\cite{cost_NOAA, 2020}. During landfall, as the strong hurricane wind field traverses inland, it propels a huge water body known as a storm surge flooding the coastal regions. The storm surge is sometimes the most destructive part of a hurricane and accounts for considerable damage~\cite{surge_web}. For instance, Hurricane Ida caused about \$55 billion in damages in Louisiana alone due to wind and storm surge damage, with additional flooding damage of about \$23 billion in the Northeastern US~\cite{IDA_NOAA}. Almost 1.2 million customers experienced power outages across eight different states. Hurricane Ian recently had a devastating impact in Florida and is expected to have incurred billions of dollars in losses, with a peak of about 2.7 million customers in a power outage~\cite{DOE_IAN}. The frequency of such high-impact, low-probability (HILP) events has increased at an alarming rate, costing about \$152.6 billion in climate-related disasters in 2021 alone in the US. Hence, there is a serious need to identify the potential impacts of these hazards on electric power systems.   

Although there have been several advancements in weather prediction models, they have not been properly utilized to analyze the potential impact of upcoming natural hazards on the power grid. Such predictive information is essential to power grid planners and operators to reduce the impacts when an event is realized~\cite{9942328, 9810633}. For instance, system operators can identify potential substations that could be inundated due to storm surges and proactively disconnect them to avoid equipment damage and facilitate fast restoration. These long-term planning strategies can help planners identify vulnerable transmission lines and propose line hardening strategies. Since hurricanes and associated storm surges have compounded spatiotemporal effects, there is an urgent need for a weather-grid impact model for outage risk assessment.    

Several existing works model the impacts of hurricanes on the power grid due to high-speed winds~\cite{7434044}. To model the spatial nature of the event, other works divided the entire power grid into zones such that each zone experiences a different wind speed~\cite{Trakas2018}. However, based on the hurricane's trajectory, each component is impacted differently as the hurricane moves inland, rendering the simplified hurricane model inaccurate for grid impact assessment. Other works model the effect of hurricanes on distribution grid~\cite{nguyen2019assessing}. Since the overall spatial exposure of a hurricane is greater than an entire distribution grid for each time step, the analysis based on the distribution system alone would not be meaningful enough.
Furthermore, none of these works model the impact of storm surges on the power grid. In~\cite{souto2022power}, a power system impact assessment framework is presented to identify potential mitigation strategies and enhance resilience against floods. A stochastic optimization framework for substation hardening against storm surge is presented in~\cite{shukla2021scenario}. \cite{Feng2022} evaluated the impacts of tropical cyclones and heatwaves on the power grid. The existing work, however, lacks the spatiotemporal impacts analysis of hurricanes characterizing the compound effects of high-speed wind and flooding. 
  
This paper aims to develop a compound spatiotemporal effect of hurricanes and storm surges on electric power systems. 
In this paper, we extend our previous work, spatiotemporal impact assessment of hurricanes~\cite{9917119}, for a more realistic assessment of hurricane impacts. First, we generate dynamic hurricane scenarios based on historical hurricane tracks and obtain storm surge scenarios based on the obtained hurricane parameters closer to the landfall. Next, sequential Monte Carlo simulation (MCS) is performed for probabilistic tracks and flooding scenarios for each time step as the hurricane moves inland. Finally, we quantify the impacts using a spatiotemporal loss metric based on the compound effect of hurricanes and floods. To the authors' knowledge, the proposed framework is the first to introduce the compound spatiotemporal impact assessment of high-speed wind and storm surges on electric power systems.        

\section{Modeling}

\subsection{Wind field Model of Hurricane}
\begin{figure}[ht]
    \centering
    \includegraphics[width=0.8\linewidth]{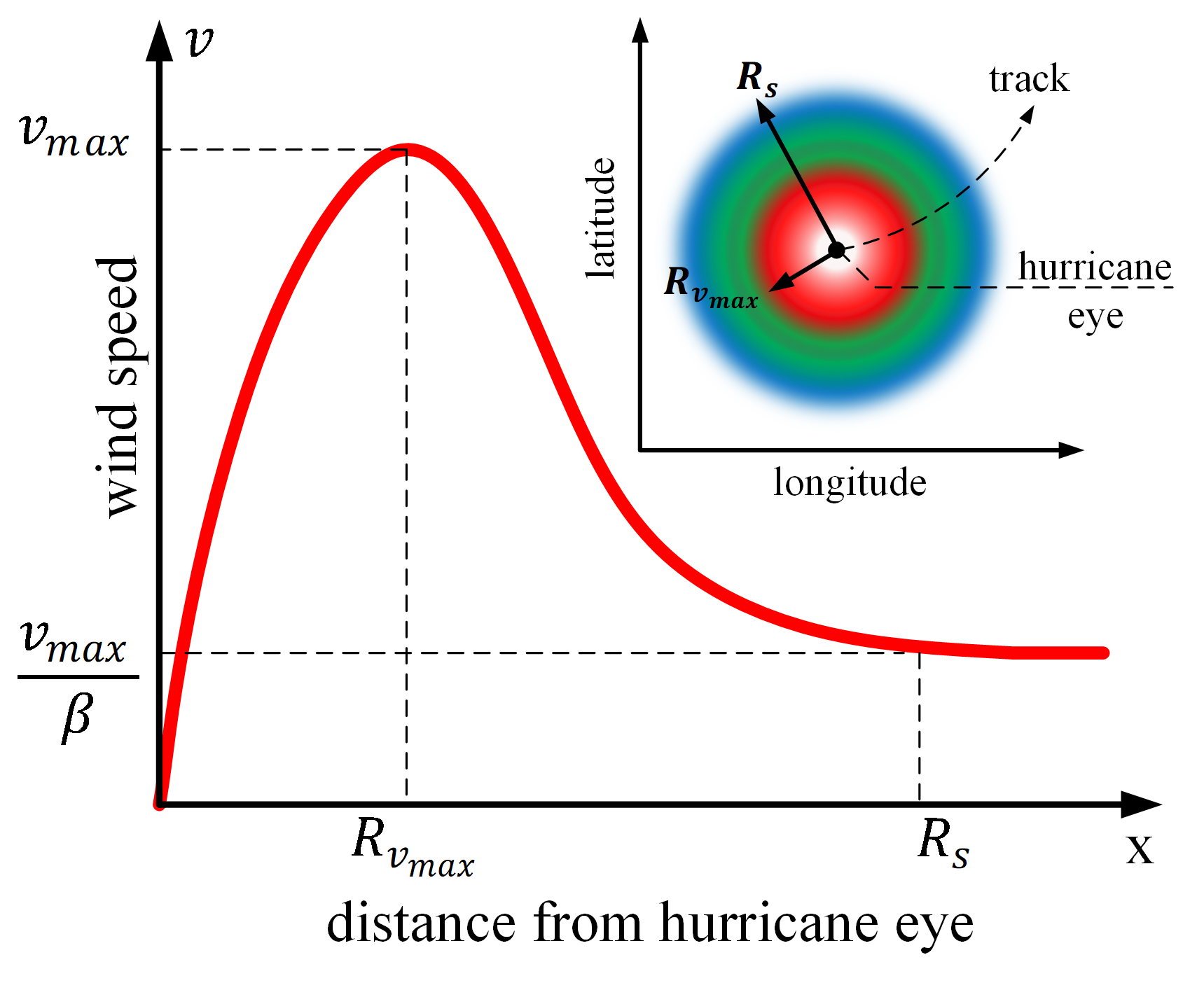}
    \caption{Static gradient wind field of a typical hurricane.}
    \label{fig:static_hurricane}
    \vspace{-5pt}
\end{figure}

A part of the impact of hurricanes originates from their wind speed. This wind speed is a vector field centered around the hurricane's eye and changes as it moves inland. This wind field can be modeled as an equation with three variables: $v_{max}$, $R_{v_{max}}$, and $R_s$~\cite{javanbakht2014risk, 9917119}. Here, $v_{max}$ measures the maximum sustained wind speed of the hurricane in knots, $R_{v_{max}}$ measures the distance to $v_{max}$ in nautical miles (nmi), and $R_s$ measures the radius of the hurricane from the hurricane's eye, also known as the radius of the outermost closest isobar (roci), in nmi. Fig.~\ref{fig:static_hurricane} demonstrates the relationship between the wind speed and the distance from the hurricane's eye, and the piecewise mathematical function that represents Fig.~\ref{fig:static_hurricane} is shown in (\ref{eq:static_eq}).
\vspace{-10pt}

\begin{equation} \label{eq:static_eq}
\small
\begin{aligned}
&v(x)= \begin{dcases} K \times v_{max}(1-exp{[-\Psi x]}) & 0 \leq x<R_{v_{max}} \\
v_{max} \exp \left[- \Lambda \left(x-R_{v_{max}}\right)\right] & R_{v_{max}} \leq x \leq R_{s} \\
0 & x>R_{s}\end{dcases} \\
&\Psi=\frac{1}{R_{v_{max}}} \ln \left(\frac{K}{K-1}\right), K>1;
\quad \Lambda = \frac{\ln \beta}{R_{s}-R_{v_{max}}}
\end{aligned}
\end{equation}

\noindent
$K$ is a known hurricane constant, $x$ is the distance from the hurricane eye, and $\beta$ is the factor that the maximum sustained wind speed will decrease at the hurricane’s boundary $(R_s)$. Using this equation, it can be assumed that the hurricane has no effect outside this boundary.

To ensure a realistic hurricane wind field, the wind parameters in (\ref{eq:static_eq}) are obtained using the International Best Track Archive for Climate Stewardship (IBTrACS). IBTrACS is a global collection of tropical cyclones that merges data from multiple agencies to create a database for historical hurricane parameters at various time steps~\cite{Knapp2010}. The hurricane wind field is generated for each time step $t$ by obtaining $\{R_{v_{max}}^t$, $R_{s}^t$, $v_{max}^t\}$ from IBTrACS and using Eq. (\ref{eq:static_eq}). 

\subsection{Storm Surge Model}
In this work, we use SLOSH (Sea, Lake, and Overland Surge from Hurricanes) model to generate probabilistic storm surge scenarios. SLOSH is a numerical storm surge model that provides surge heights around the coastal regions from historical or hypothetical hurricanes based on parameters such as atmospheric pressure, hurricane tracks, forward speed, and so forth~\cite{glahn2009role}. It was developed by National Oceanic and Atmospheric Administration (NOAA) and is currently being used by several federal agencies, including the national hurricane center (NHC) and the federal emergency management agency (FEMA), for flood advisories and evacuation. Furthermore, NHC provides a SLOSH Display Package (SDP) tool in which users can generate flooding scenarios based on the direction, forward speed, and intensity of the hurricane followed by the sea tide level~\cite{SDP}. The provided surge heights are above the elevation referenced in the North American Vertical Datum of 1988 (NAV88). The tool also has an added functionality to subtract land elevations so that the surge level is referenced above the ground level. Details on the SDP tool and other surge-related products from NOAA can be found at~\cite{SDP}. 

SDP contains several coastal basins on which surge scenarios can be created. Furthermore, SDP provides two different surge analyses based on hurricane simulations: Maximum Envelope of Water (MEOW) and Maximum of the MEOWs (MOM). The MEOWs reflect the worst-case snapshots of the storm surge for hurricanes of particular intensity and forward direction but with different landfall locations. This work uses MEOW for storm surge scenarios to identify potential flooding locations. The MEOW scenarios from SDP provide the inundation level for each defined grid in a basin. Let $\mathcal{X}_\mathcal{S}$ be the geographical coordinate of a substation $\mathcal{S}$. Then, we define $\mathcal{X}^\mathcal{B}_{\mathcal{S}, h}$ as the inundation level, $h$, for substation $\mathcal{S}$ situated at $\mathcal{X}$ for basin $\mathcal{B}$. Since the inundation level is spatially distributed, the expected value of inundation around 0.5 miles of $\mathcal{X}_\mathcal{S}$ is assumed to be $\mathcal{X}^\mathcal{B}_{\mathcal{S}, h}$.

\subsection{Impact Assessment Model}
Combined, the hurricane and storm surge creates a spatiotemporal effect on the power grid. The spatiotemporal impact of hurricanes on the power grid is generally guided by the fragility curves of the transmission lines~\cite{7801854}. Let $\Gamma_{l}^{t,\zeta}$ be the maximum wind speed on line $l$ at time step $t$ due to a hurricane traversing in track $\zeta$. Then the outage probability of line $l$ at time step $t$ due to a hurricane traversing in track $\zeta$ is given by (\ref{eq:hurricane_outage})      
\vspace{-1em}
\begin{equation}
\small
\begin{aligned}
&\mathbb{P}_{out}^{t,\zeta}(l)= \begin{dcases} 0 & \Gamma_{l}^{t,\zeta} < v_{cri} \\
\frac{\Gamma_{l}^{t,\zeta} - v_{cri}}{v_{col} - v_{cri}}  & v_{cri} \leq \Gamma_{l}^{t,\zeta} < v_{col} \\
1 & \Gamma_{l}^{t,\zeta} \geq v_{col} \end{dcases} \\
\end{aligned}
\label{eq:hurricane_outage}
\end{equation}

\noindent
where $v_{cri}$ = 48.59 knots and $v_{col}$ = 106.91 knots are the critical wind speed beyond which a transmission line is affected by the hurricane and the wind speed at which the transmission line collapses, respectively~\cite{9917119}. If $\delta_{l}^{t,\zeta} \in \{0,1\}$ denotes the line status of line $l$ at time $t$ for hurricane in track $\zeta$, then $\delta_{l}^{t + 1,\zeta} \leq  \delta_{l}^{t,\zeta}$. This ensures that if a line $l$ experiences an outage at time $t$, it remains out of service for the remaining duration of the hurricane.    

The impact assessment from storm surges can be modeled similarly by defining the outage probability of substations. Although FEMA has provided a general fragility level for substations in their HAZUS tool~\cite{hazus_flood}, we relax the curve through Weibull stretched exponential function as the substations around coasts have some hardening measures already in place. Hence, the outage probability of a substation $\mathcal{S}$ for a given inundation level $h$ at each basin $\mathcal{B}$ is given by (\ref{eq:flood_outage})

\vspace{-1em}
\begin{equation} 
\small
\mathbb{P}^{\mathcal{S}}_{out}(\mathcal{X}^\mathcal{B}_{\mathcal{S}, h}) = 1 - exp\left[- {\left(\frac{\mathcal{X}^\mathcal{B}_{\mathcal{S}, h}}{a}\right)^b}\right] \\
\label{eq:flood_outage}
\end{equation}

\noindent
where $a\in \mathbb{R}^+$ and $b>2$ are known constants and determine the shape of the fragility curve. When a substation is flooded and deemed out-of-service, all transmission lines connected to and from the substation are disconnected. If $\delta_{l,\mathcal{S}}^{t,\zeta} \in \{0,1\}$ denotes the line status of line $l$ connected to substation $\mathcal{S}$ at time $t$ for hurricane in track $\zeta$, then $\delta_{l, \mathcal{S}}^{t + 1,\zeta} \leq  \delta_{l,\mathcal{S}}^{t,\zeta}$. This ensures that if a line $l$ experiences an outage at time $t$ due to a storm surge, it remains out of service for the remaining event duration. 
When simulating storm surges as an effect of hurricanes, it is important to avoid the redundancy of impact that both transmission line and substation outages can have on the power grid. Let $\mathcal{L}^t_\mathcal{H}$ be the set of lines affected by hurricanes and $\mathcal{L}^t_\mathcal{S}$ be the set of lines out of service due to a flooded substation at time $t$. Then, the total number of lines affected by the compound effect of the hurricane and substations inundation due to storm surge is given by $\mathcal{L}^t_\mathcal{H} \cup \mathcal{L}^t_\mathcal{S}$.    

\vspace{-0.5em}
\section{Overall Framework}
\begin{figure}[t]
    \centering
    \includegraphics[width=0.85\linewidth]{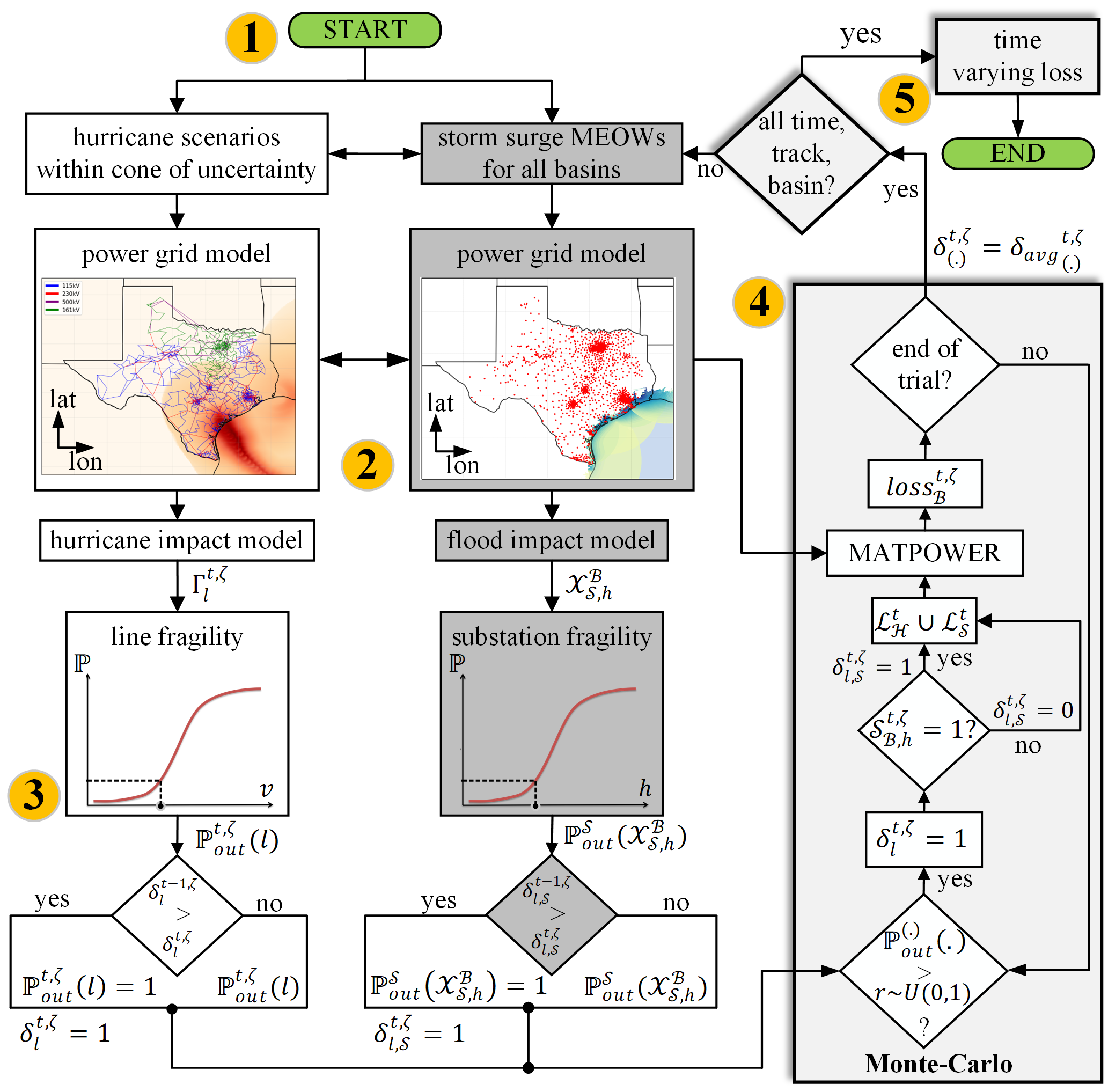}
   \caption{Overall architecture for generating hurricane and storm surge scenarios and grid impact assessment.}
    \label{fig:overall_architecture}
    \vspace{-5pt}
\end{figure}

This section describes the overall approach to assess the spatiotemporal impact of hurricane and storm surges on the electric power systems; see Fig.~\ref{fig:overall_architecture}. First, synthetic hurricane tracks were generated based on historical data from IBTrACS. The hurricane impact model is then used to obtain the wind speed experienced by each transmission line. Similarly, flooding scenarios are obtained from SLOSH basins for the given characteristic of the hurricane. The fragility function defined in (\ref{eq:hurricane_outage}) and (\ref{eq:flood_outage}) provides the outage probability of transmission lines due to hurricanes and the outage probability of substations due to coastal flood inundation. Finally, Monte Carlo simulations (MCS) are conducted to obtain the probabilistic loss for the system. All flood scenarios are considered equally likely. The final spatiotemporal system-level loss metric for each time step is the obtained. The loss metric is the expected value of the loss obtained for the hurricane track and the flood basin for that particular time step. The following subsections detail the overall framework. 

\subsection{Generating Hurricane and Storm Surge scenarios}
\begin{figure}[t]
    \centering
        \includegraphics[width=0.8\linewidth]{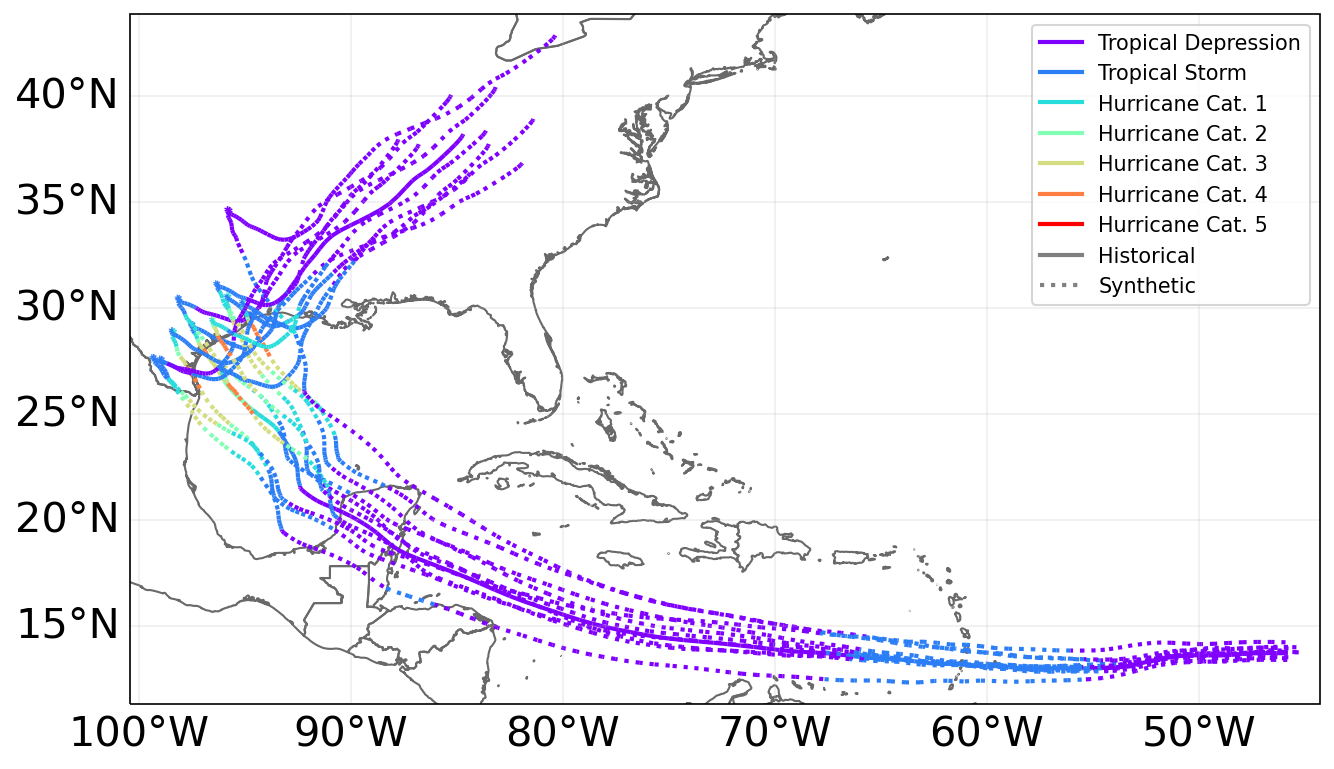}
        \caption{Hurricane Harvey track and other synthetic tracks obtained from CLIMADA~\cite{Climada}.}
        \label{fig:static_harvey}
        \vspace{-15pt}
\end{figure}

The geographical data and parameters are obtained for Hurricane Harvey, that hit the coast of Texas in August 2017 using IBTrACS. Synthetic hurricane tracks are obtained by perturbing several characteristics of Hurricane Harvey, such as a shift in the initial point of origin, the amplitude of translational speed, bearing angle, landfall decay, etc. The synthetic tracks are obtained from the Climate adaptation (CLIMADA) tool. CLIMADA is a probabilistic natural catastrophe damage model that facilitates climate-based simulations~\cite{Climada}. CLIMADA can generate synthetic tracks in IBTrACS format based on the above perturbations. To synchronize the time steps for the entire duration of the hurricanes, the data is observed every two hours, and the hurricane wind field is generated for each $t$ using (\ref{eq:static_eq}). Fig.~\ref{fig:static_harvey} shows the synthetic tracks generated from CLIMADA, and the track with a solid line represents the actual track of Hurricane Harvey from IBTrACS. NOAA predicts that 7 out of 10 times, the estimated hurricane track is within the cone of uncertainty~\cite{NOAA_uncertainty}. Hence, five synthetic tracks are selected to be within the cone of uncertainty when forecasted over 120 hours of landfall, i.e., within 240 nmi as per~\cite{NOAA_uncertainty}.    

For storm surge assessment, five different Texas basins are selected; $\mathcal{B} \in$ \{Matagorda, Corpus, Galveston, Laguna, Sabine\}. MEOW scenarios are obtained for all five basins for category 4 hurricanes moving in the northwest direction with a translational speed of 10-15 miles per hour and mean to high tide conditions. One major drawback of using MEOW maps is they are snapshots of the storm surge and carry no time information. However, we can assume that the components are not inundated until the hurricane approaches. Hence, we assign an activation flag for each component per basin at each time step. This is discussed in the next subsection. 

\subsection{Power Grid Impact Assessment using MCS}
\begin{figure}[t]
    \centering
    \begin{subfigure}[t]{0.8\linewidth}
        \centering
        \includegraphics[width=0.8\linewidth]{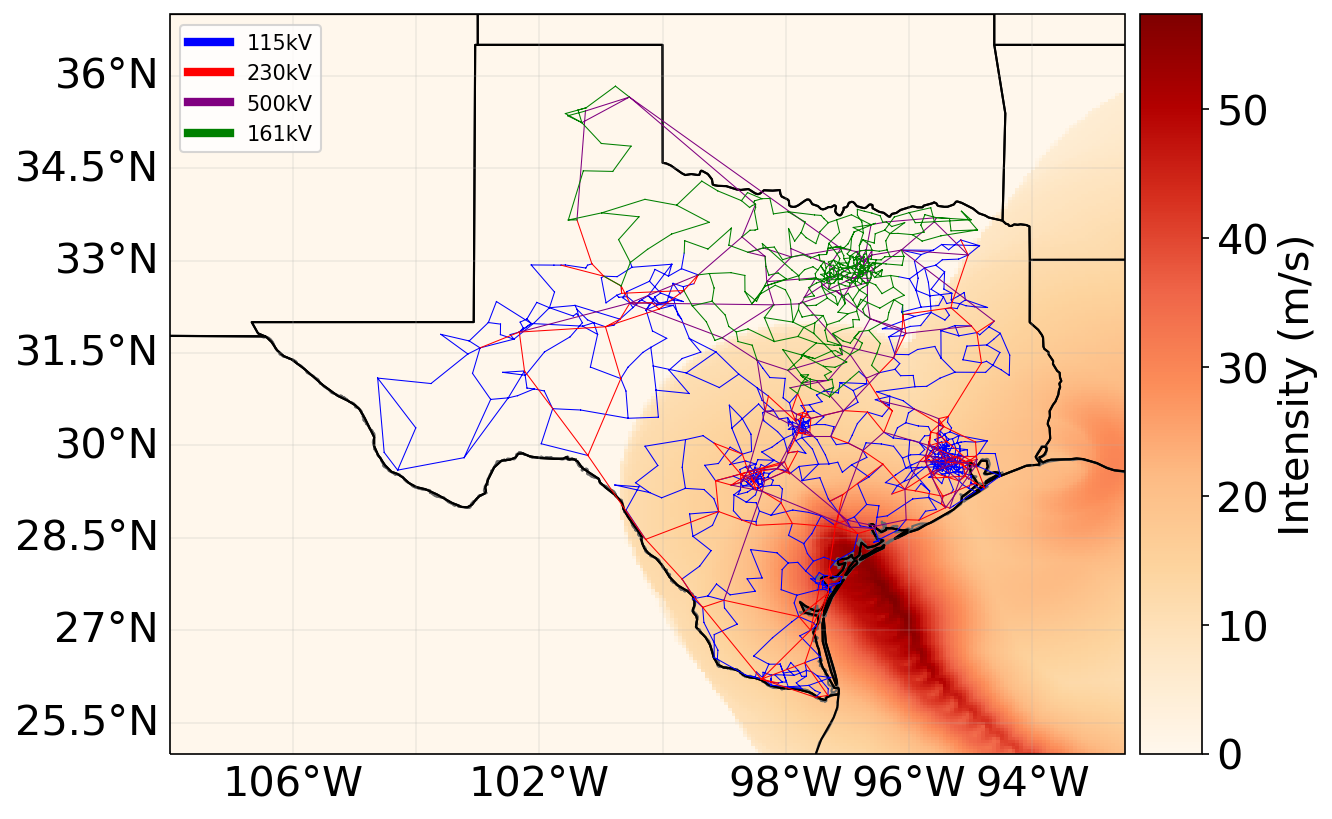}
        \caption{}
        \label{fig:dynamic_harvey}
    \end{subfigure}
    \begin{subfigure}[t]{0.8\linewidth}
        \centering
        \includegraphics[width=0.8\linewidth]{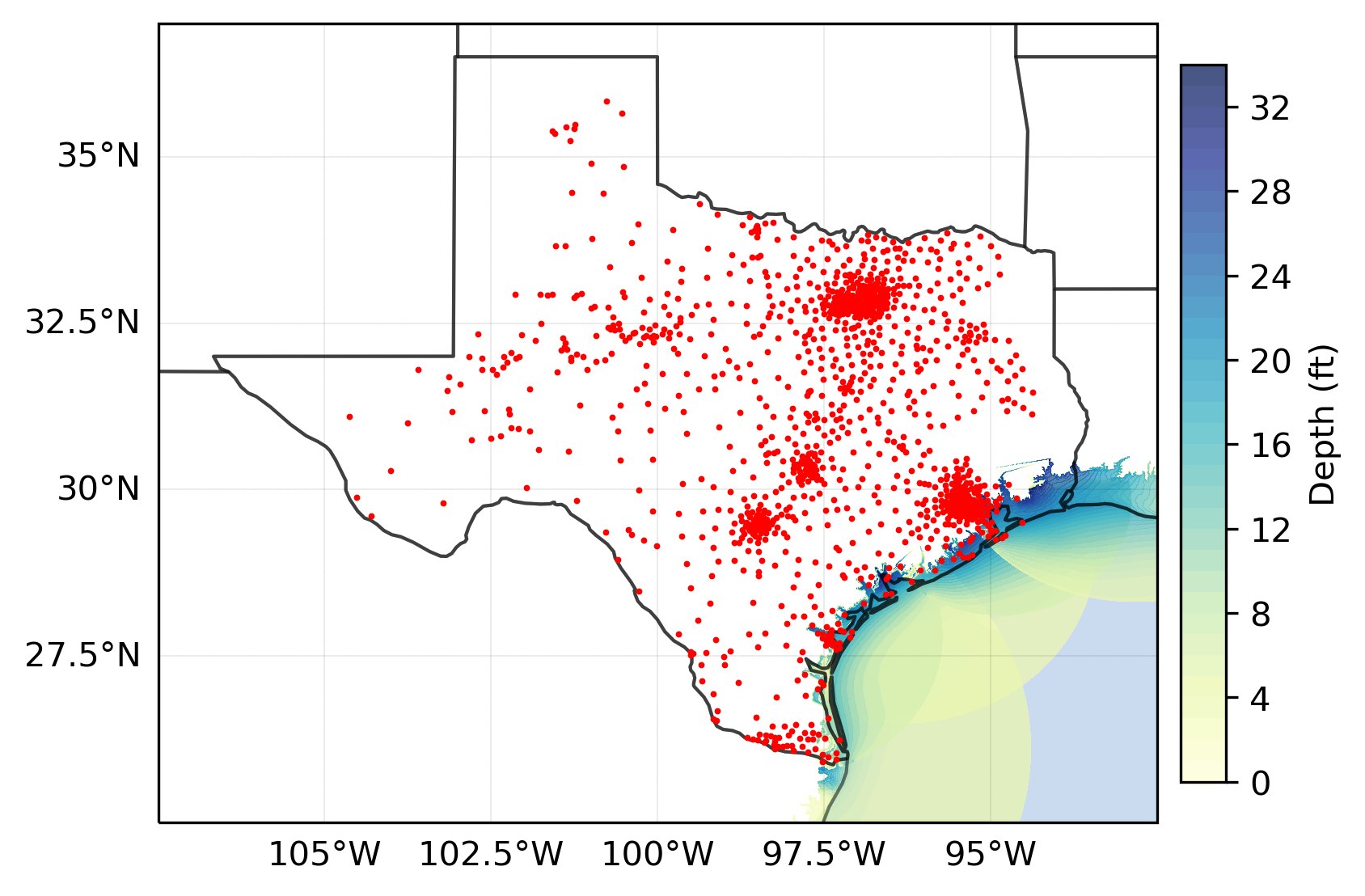}
        \caption{}
        \label{fig:flood_scenario}
    \end{subfigure}
    \caption{a) Wind field of Hurricane Harvey and b) Substation flooding scenario, above ground level, for Texas basins.} 
    \label{fig:hurricane_flood_scenario}
    \vspace{-0.5em}
\end{figure}

This work uses a synthetic 2000-bus Texas power grid model to identify the compound probabilistic impact of hurricane and storm surges on the power grid~\cite{7725528}. The grid model maps the synthetic buses, substations, and transmission lines on the geographical footprint of Texas. There are 1250 substations and 1918 transmission lines at 4 different voltage levels --- 115 kV, 161 kV, 230 kV, and 500 kV. The total generation capacity of the system is 96291.5 megawatts (MW), and there are 1125 load units with a total size of 67109.2 MW. When the hurricane moves inland, $\Gamma_{l}^{t,\zeta}$ is obtained for each $l$ and $\mathcal{X}^\mathcal{B}_{\mathcal{S}, h}$ is obtained for each $\mathcal{S}$ from the MEOW maps. It is assumed that all substations at the coast are elevated at a level of 3 ft from the ground level. Fig.~\ref{fig:hurricane_flood_scenario} shows the wind field of Hurricane Harvey and MEOW maps on five different basins on the footprint of Texas.     

For each $\zeta$, $\mathcal{B}$, and $t$, MCS is performed for several trials. For each trial, $\mathbb{P}_{out}^{t,\zeta}(l)$ and $\mathbb{P}^{\mathcal{S}}_{out}(\mathcal{X}^\mathcal{B}_{\mathcal{S}, h})$ is compared with a uniform random number, $r\sim U(0,1)$, to identify $\mathcal{L}^t_\mathcal{H} \cup \mathcal{L}^t_\mathcal{S}$. As discussed before, MEOW maps are time-independent. However, we define an activation flag $\mathcal{S}^{t, \zeta}_{\mathcal{B}, h} \in \{0,1\}$ such that $\mathcal{S}^{t, \zeta}_{\mathcal{B},h} = 1$ makes the substation $\mathcal{S}$ inundated with depth $h$ above the ground elevation for hurricane track $\zeta$, SLOSH basin $\mathcal{B}$, and time $t$. In this work, we do not model any drainage system, and hence it is assumed that for future time steps $t+1$ the inundation level is the same. The objective here is to identify the time stamp when the substation gets flooded to analyze the compound spatiotemporal impact of two different hazards at that time step. The information, $\mathcal{L}^t_\mathcal{H} \cup \mathcal{L}^t_\mathcal{S}$, is then sent to the power grid simulator to remove the respective branches from the system. Finally, the loss for the particular $t$ is observed as the total load disconnected from the main grid. The MCS trial is conducted until the loss converges to some value.

We assume that the inundation scenarios for each $\mathcal{B}$ and each of the hurricane tracks $\zeta$ are equally likely. If $N_\zeta$ is the total number of tracks under consideration and $N_\mathcal{B}$ is the total number of basins, then the system level load loss at each time step $t$ is given by (\ref{eq:final_loss})

\vspace{-1.5em}

\begin{equation}
\small
    loss_t = \frac{1}{N_\zeta \times N_\mathcal{B}}\sum_{\zeta = 1}^{N_\zeta}{\sum_{\mathcal{B}} {loss_{\mathcal{B}}^{t,\zeta}}}
    \label{eq:final_loss}
\end{equation}

\vspace{-1em}
\section{Results and Analysis}
The hurricane data is obtained from IBTrACS~\cite{Knapp2010}, and synthetic hurricanes are generated using CLIMADA~\cite{Climada} in Python. SDP~\cite{SDP} is used to obtain the MEOW maps for Texas basins. The power grid is modeled in MATPOWER 7.1~\cite{5491276}, and MCS on the power grid model is conducted in MATLAB R2021a.  

\begin{figure}[t]
    \centering
    \begin{subfigure}[t]{0.235\textwidth}
        \centering
        \includegraphics[width=\linewidth]{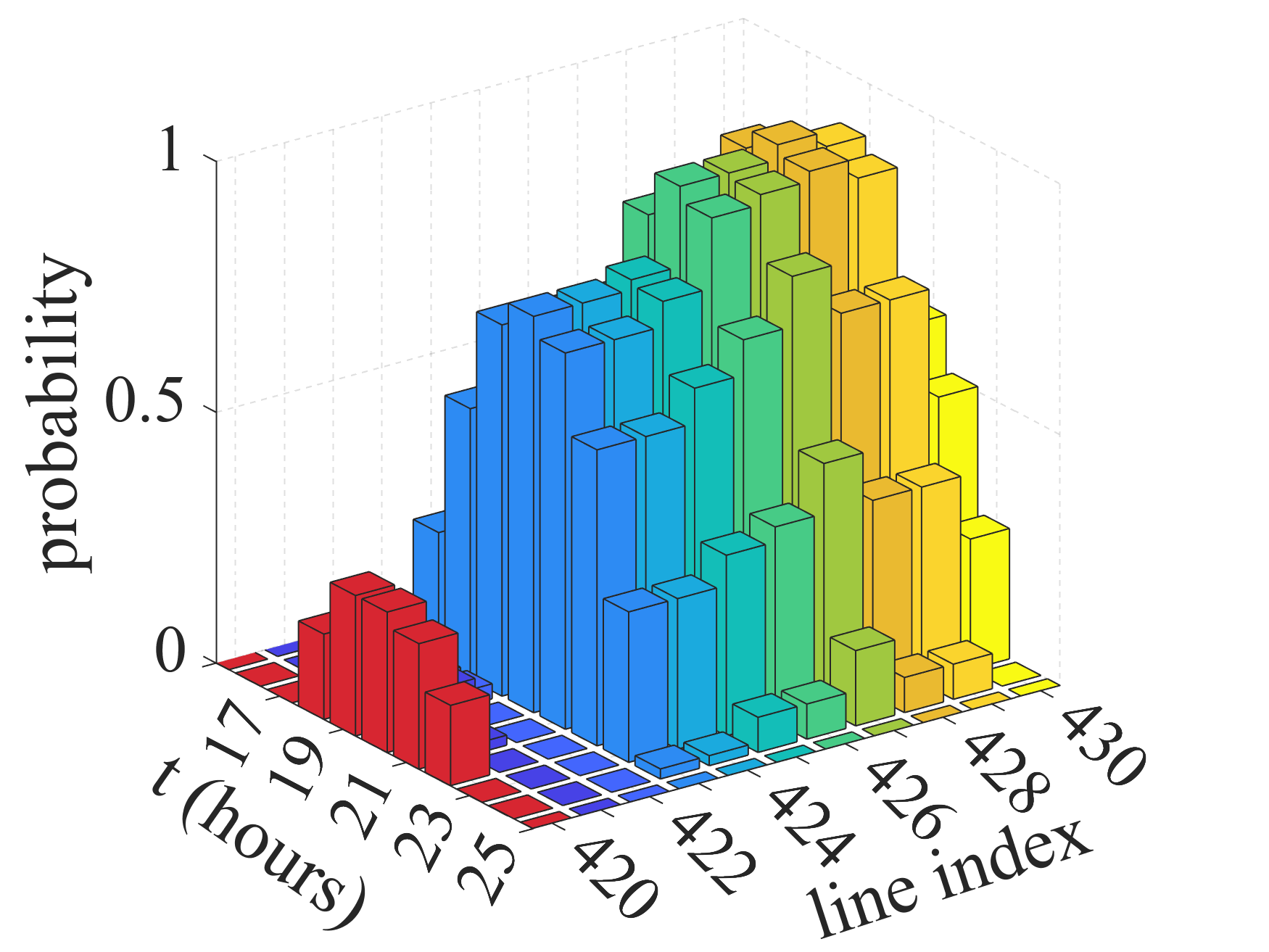}
        \caption{}
        \label{fig:line_outage_probability}    
    \end{subfigure}
    \hspace*{\fill}
    \begin{subfigure}[t]{0.235\textwidth}
        \centering
        \includegraphics[width=\linewidth]{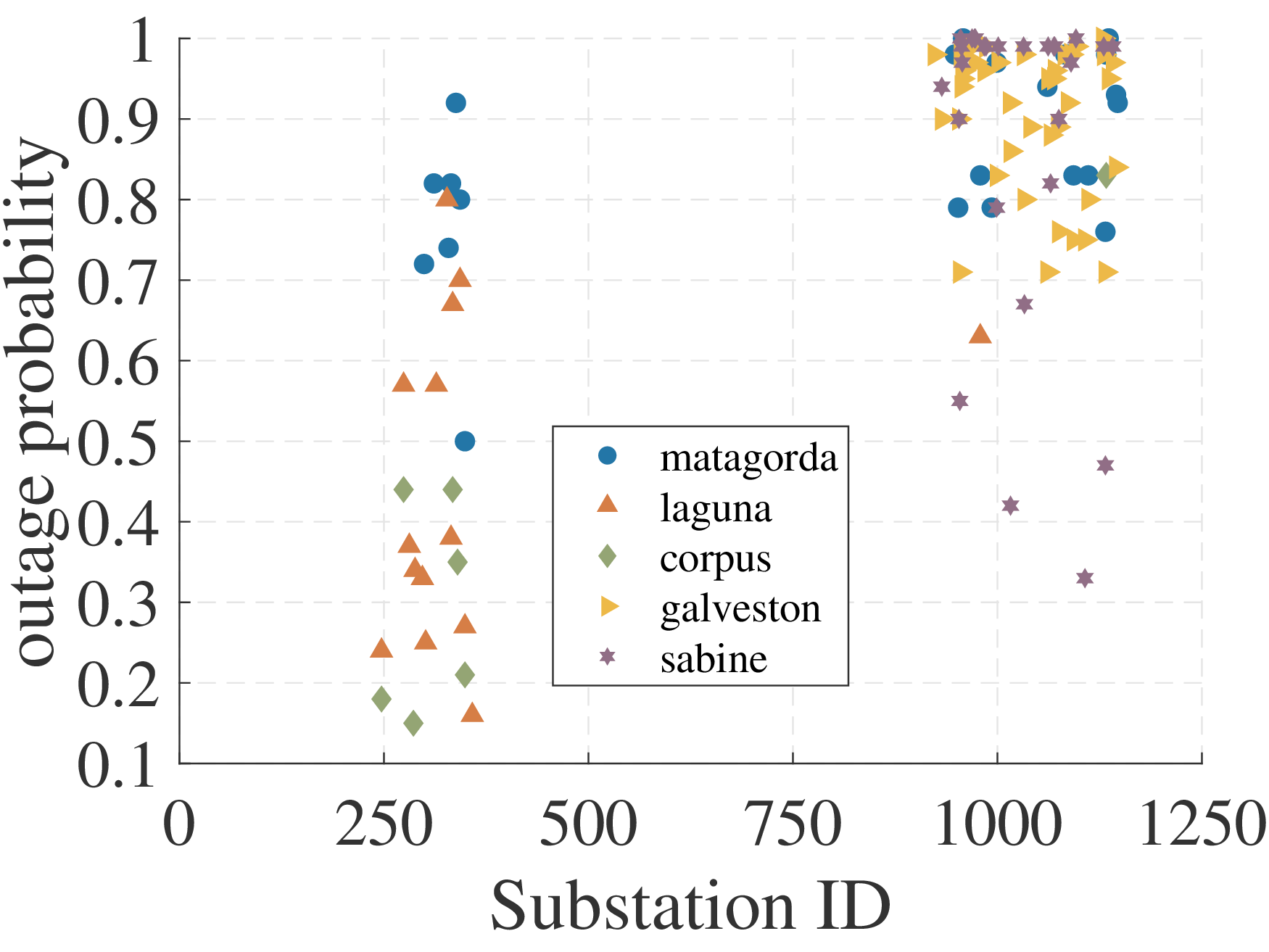}
        \caption{}
        \label{fig:substation_outage_probability}
    \end{subfigure}
    \caption{a) Line outage probability for selected lines and the original Harvey's track. b) Substation outage probability for all storm surge scenarios.} 
    \label{fig:hurricane_flood_outages}
    \vspace{-1.5em}
\end{figure}

From the IBTrACS, the hurricane's initial location is in the North Atlantic Ocean, which is observed at $``2017-08-16~06:00:00''$ and hence creates no impact on the power grid. Thus, the simulation for the impact assessment begins at the $100^{th}$ time step, i.e., at the advisory of $``2017-08-24~12:00:00''$. To avoid any confusion, this timestamp is referred to as $t=0$ hours for the rest of the paper. Fig.~\ref{fig:line_outage_probability} shows the outage probability of a set of transmission lines due to the original Hurricane Harvey track. It can be observed that with the moving nature and intensity of the hurricane, the outage probability of each line changes. The probability is maximum when the line experiences wind speed closer to $v_{col}$ and gradually decreases as the hurricane decays while moving inland. Fig.~\ref{fig:substation_outage_probability} represents the outage probability of substations for five different Texas basins. It is seen that the substation's vulnerability depends on the location and basin.

\begin{figure}
    \centering
    \includegraphics[width=\linewidth]{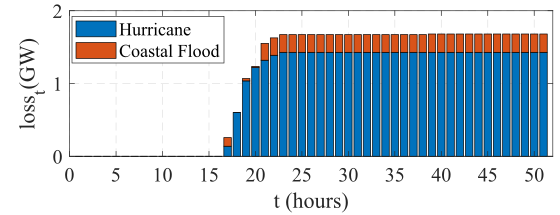}
    \caption{Overall loss comparison between the impact of hurricane alone and the compound impact of the hurricane and coastal flood for $\zeta = 1$.}
    \label{fig:hurricane_flood_compound}
    \vspace{-10pt}
\end{figure}

It was found that 800 Monte Carlo trials are enough to achieve convergence for any simulation case. Hence, for each $\zeta$, $\mathcal{B}$, and $t$, 800 Monte Carlo trials are conducted. Fig.~\ref{fig:hurricane_flood_compound} shows the comparison of losses when considering the impact of the hurricane alone with the compound effect of the hurricane and the coastal flood for $\zeta=1$. It can be observed that considering substation flooding scenarios incur an additional loss of $250~MW$ by $t=50$ hours due to substation outages.

\begin{figure}[t]
    \centering
    \includegraphics[width=0.8\linewidth]{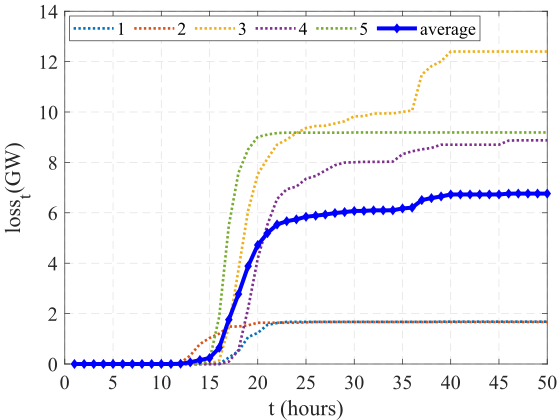}
    \caption{Time-varying loss for each hurricane track. For each track, the loss at each time step also includes the expected value of loss over entire flood basins.}
    \label{fig:final_losses}
    \vspace{-1em}
\end{figure}

Fig.~\ref{fig:final_losses} represents the spatiotemporal loss for the compound impact of hurricane and storm surge. The dashed curves represent the loss for all basins for each track, whereas the solid blue curve represents $loss_t$ obtained from (\ref{eq:final_loss}). The loss is not incurred until $t=12$ for any $\zeta$ under consideration, and the increase is significant once the hurricane moves inland, followed by the compounded impact of storm surge. The loss increases from $loss_{12} = 9.4072$ MW to $loss_{22} = 5536.1$ MW before saturating at $loss_{22} = 6761.97$ MW as the intensity of the extreme events decreases.  

\vspace{-0.5em}
\section{Conclusion}
This paper presents the compounded spatiotemporal impact assessment of hurricanes and storm surges on the power grid. A spatiotemporal loss metric is identified based on a probabilistic model. Simulations showed that storm surges could flood the coastal substations to incur an additional loss in the system. The hurricane intensity decays, and the overall loss saturates at some point, after which system restoration is required to bring back the power to unserved parts of the grid. The proposed method can help system operators identify vulnerable regions and propose hardening strategies to either withstand part of the damage or facilitate quick restoration. Furthermore, such a framework can be beneficial in identifying proactive planning strategies. Another possible direction for future research would develop feasible restoration plans to enhance the grid's resilience following the aftermath.    

\vspace{-0.2em}
\bibliographystyle{IEEEtran}
\bibliography{ref.bib}
\end{document}